\documentclass[onecolumn,aps,prd,preprintnumbers,showpacs,superscriptaddress,nofootinbib,amsmath,amssymb,floats,floatfix,showkeys,notitlepage,longbibliography]{revtex4-1}

\usepackage[utf8]{inputenc}
\usepackage[english]{babel}
\usepackage{graphicx}
\usepackage{dcolumn}
\usepackage[dvipsnames]{xcolor}
\usepackage[T1]{fontenc}

\usepackage{mathrsfs}  
\usepackage{cases}
\usepackage{bm}
\usepackage{academicons}
\usepackage{mathtools, nccmath}
\usepackage{fancyhdr}
\usepackage{tikz,xcolor}

\usepackage{tensor}
\usepackage{orcidlink}
\usepackage[normalem]{ulem}
\usepackage{lipsum}
\usepackage{soul}
\usepackage{cancel}
\usepackage{stackengine,scalerel}
\usepackage{mathabx}
\usepackage{hyperref}
\usepackage{tabularx}
\hypersetup{colorlinks, linkcolor={red},citecolor={blue},urlcolor={blue}}

\definecolor{lime}{HTML}{A6CE39}
\DeclareRobustCommand{\orcidicon}{
	\begin{tikzpicture}
	\draw[lime, fill=lime] (0,0) 
	circle [radius=0.16] 
	node[white] {{\fontfamily{qag}\selectfont \tiny ID}};
	\draw[white, fill=white] (-0.0625,0.095) 
	circle [radius=0.007];
	\end{tikzpicture}
	\hspace{-2mm}
}

\fancypagestyle{plain}{%
  \fancyhf{}
  \fancyfoot[C]{\iffloatpage{}{\thepage}}
  }
\foreach \x in {A, ..., Z}{%
	\expandafter\xdef\csname orcid\x\endcsname{\noexpand\href{https://orcid.org/\csname orcidauthor\x\endcsname}{\noexpand\orcidicon}}
}

\begin{document}

\title{Regular Black Hole Formation and Gamma-Ray Burst from Matter Conversion}

\author{Vitalii Vertogradov
\orcidlink{0000-0002-5096-7696}}
\email{vdvertogradov@gmail.com}
\affiliation{Wilczek Quantum Center, Shanghai Institute for Advanced Studies, Shanghai 201315, China}
\affiliation{University of Science and Technology of China, Hefei 230026, China}
\affiliation{Physics department, Herzen state Pedagogical University of Russia,
48 Moika Emb., Saint Petersburg 191186, Russia.}
\affiliation{Center for Theoretical Physics, Khazar University, 41 Mehseti Street, Baku, AZ-1096, Azerbaijan.}
\affiliation{SPB branch of SAO RAS, 65 Pulkovskoe Rd, Saint Petersburg
196140, Russia.}
\author{Zhanna Kuznetsova 
\orcidlink{0009-0003-1893-1495}}
\email{kuznetsovaaj133@gmail.com}
\affiliation{Physics department, Herzen state Pedagogical University of Russia,
48 Moika Emb., Saint Petersburg 191186, Russia.}
\author{Yu Shi
\orcidlink{0000-0002-5096-7696}}
\email{yu_shi@ustc.edu.cn}
\affiliation{Wilczek Quantum Center, Shanghai Institute for Advanced Studies, Shanghai 201315, China}
\affiliation{University of Science and Technology of China, Hefei 230026, China}

\begin{abstract}
During the gravitational collapse of a massive star into a regular black hole, a new form of matter must be produced in order to prevent the formation of a central singularity. Since such matter is not present in the initial stellar configuration, it must emerge dynamically during the collapse. This formation process is expected to be accompanied by a strong release of energy in the form of electromagnetic radiation, which may be observable. Here we investigate the gravitational collapse of baryonic matter into Dymnikova-Hayward-Bardeen regular black holes. We estimate the radiation density and the corresponding bolometric luminosity generated by the formation of the matter sector responsible for singularity avoidance. We show that such processes   provide a possible mechanism for gamma-ray bursts. Moreover, compatibility with gamma-ray burst requires small regularization effects. As a result, the corresponding regular black holes differ weakly from the Schwarzschild black hole.
\end{abstract}

\date{\today}



\maketitle
\section{Introduction}

In 1916, Karl Schwarzschild found the first exact vacuum solution of the Einstein equations. This solution, now known as the Schwarzschild metric, remains one of the most important spacetimes in general relativity and provides the standard qualitative description of many effects associated with black holes. However, the Schwarzschild geometry contains a fundamental pathology: in the limit \(r\to0\), curvature invariants diverge, indicating the presence of a spacetime singularity. The appearance of such an infinity signals that classical general relativity cannot provide a complete description of the innermost region of a black hole.

The singular nature of gravitational collapse is not merely a peculiarity of the Schwarzschild solution. Penrose proved his celebrated singularity theorem, according to which, under appropriate causal assumptions and energy conditions, the formation of a trapped surface in gravitational collapse inevitably leads to geodesic incompleteness~\cite{bib:penrose}. Thus avoiding the formation of a singularity requires that at least one of the assumptions of the theorem be violated. In the context of regular black holes, the most physically relevant possibility is the violation of the strong energy condition, which expresses the attractive character of gravity for ordinary matter.

Such a violation is not alien to modern physics. On cosmological scales, the accelerated expansion of the Universe is commonly attributed to a dark-energy component, which violates the strong energy condition. This observation motivates the idea that, at extremely high densities, gravitational collapse may also enter a regime where the effective matter source ceases to behave as ordinary baryonic matter. In the 1960s, Bardeen proposed the first phenomenological black-hole solution without a central singularity~\cite{bib:bardeen}. Such objects are now known as regular black holes. In order to evade the conclusion of the Penrose theorem, regular black holes must contain matter that violates the strong energy condition in the deep interior~\cite{bib:zaslavskiy}.

A closely related physical idea was proposed by Gliner~\cite{bib:gliner} and Sakharov~\cite{bib:sakharov}. They suggested that, at sufficiently high densities, matter may undergo a transition into a vacuum-like state. In the context of gravitational collapse, this means that the central region of the collapsing object may approach an equation of state of the form
\begin{equation}
P=-\rho,
\end{equation}
so that the innermost geometry tends to a de Sitter spacetime. For this reason, the regular central region is usually referred to as a de Sitter core.

The ideas of Gliner and Sakharov have inspired many models of regular black holes. A large class of such solutions has been constructed in different physical settings, including phenomenological de Sitter-core models, nonlinear electrodynamics, anisotropic fluids, and effective matter sectors motivated by quantum-gravity corrections~\cite{bib:hay,bib:dym1,bib:dym2,bib:khlopov1,bib:khlopov2,bib:khlopov3,bib:khlopov4,bib:khlopov5,bib:ali2025cqg,bib:maxim2024podu,bib:emmanuele,bib:yi2026epjc}.

There is, however, a crucial observational problem. Once a black hole has formed, an external observer has no direct access to its central region. Therefore, it is extremely difficult to determine whether the interior contains a singularity or a regular core. A large number of works have studied shadows, quasi-normal modes, and thermodynamic properties of regular black holes~\cite{bib:shakun2024pcs,bib:ali2026plb,bib:shadow1,bib:shadow2,bib:shadow3,bib:shadow4,bib:shadow5}. These studies show that, for suitable values of the regularization parameters, regular black holes can be made compatible with the observational constraints obtained by the Event Horizon Telescope~\cite{bib:eht1,bib:eht2}. However, similar agreement with present shadow data can also be achieved for singular black holes. Hence, shadow observations alone may not be sufficient to determine whether the black-hole interior is singular or regular.

For this reason, one has to look not only at the final compact object, but also at the process of its formation. Regular black holes require a matter sector that violates the strong energy condition. Such a sector is not present in ordinary massive stars. Therefore, if a regular black hole is formed in gravitational collapse, the matter responsible for singularity avoidance must be generated dynamically during the collapse. At critical densities, ordinary stellar matter may undergo a phase transition into a new state, which prevents the formation of the central singularity~\cite{bib:vertogradov2025podu,bib:ali2025jcap,bib:vertogradov2025epjc,bib:shatov2025cpc}. If this transition occurs before the event horizon completely hides the relevant region, part of the released energy may escape in the form of electromagnetic radiation.

Gamma-ray bursts (GRBs) provide a natural observational context for such a mechanism. GRBs were first identified as cosmic high-energy transients by Klebesadel, Strong and Olson~\cite{bib:klebesadel1973}. They are now known to be among the most luminous electromagnetic explosions in the Universe. The observed population is traditionally divided into short and long bursts, with with the threshold set at at approximately two seconds~\cite{bib:kouveliotou1993}. Long GRBs are usually associated with the collapse of massive stars and the formation of a relativistic jet powered by a compact central engine~\cite{bib:woosley1993,bib:macfadyen1999}. The association of long GRBs with core-collapse supernovae was strongly supported by the detection of SN 1998bw in connection with GRB 980425~\cite{bib:galama1998}. Short GRBs, in contrast, are usually connected with compact binary mergers, a picture confirmed by the multi-messenger event GW170817/GRB 170817A~\cite{bib:abbott2017gw170817,bib:goldstein2017grb170817a}.

Despite this general classification, the physical origin of GRBs is not completely understood. The central engine, the jet-launching mechanism, the baryon loading of the jet, the prompt-emission mechanism, and the diversity of observed durations remain open problems. In particular, several recent events challenge the standard picture. The burst GRB 221009A, often called the ``Brightest Of All Time'', was the brightest GRB detected in terms of prompt peak flux and fluence and reached an isotropic-equivalent energy of order \(10^{55}\,{\rm erg}\)~\cite{bib:burns2023boat,bib:frederiks2023grb221009a}. Even more recently, GRB 250702B was observed as an exceptionally long and repeating high-energy transient, with activity lasting for a day and with energetics of at least \(E_{\gamma,\rm iso}\sim10^{54}\,{\rm erg}\)~\cite{bib:levan2025grb250702b,bib:gompertz2025grb250702b}. Events of this type indicate that the known GRB classes may not exhaust all possible mechanisms of high-energy transient emission. This motivates the search for additional channels capable of producing large electromagnetic luminosities during compact-object formation.

In the present work, we investigate the gravitational collapse of baryonic matter leading to the formation of Dymnikova-Hayward-Bardeen regular black holes. During the transition of baryonic matter into a new type of matter that prevents the formation of a singularity, energy is released. We estimate the density of this radiation and the corresponding luminosity. We show that such a process can produce luminosities comparable to those of gamma-ray bursts. This requires the regularization effects near the outer horizon to be weak of order $\sim10^{-1}-10^{-2}.$
Consequently, the resulting regular black holes differ only slightly from the Schwarzschild geometry near the outer horizon. We argue that a phase transition during gravitational collapse may therefore serve not only as a possible source of gamma-ray-burst-like emission, but also as a way to constrain the regularization parameters of the final compact object. In combination with future shadow observations, such constraints may help distinguish regular black holes from singular ones.

The paper is organized as follows. In Sec.~II, we discuss a model of gravitational collapse of baryonic matter into regular black holes and derive the radiation density generated during the phase transition. In Sec.~III, we construct a smooth matching between the Kiselev metric and the Dymnikova-Hayward-Bardeen regular black-hole metrics. This allows us to determine the critical radius at which the radiation vanishes, i.e., the radius where baryonic matter has completely transformed into the new matter sector. In Sec.~IV, we analyze the radiation density released in the form of electromagnetic waves during the phase transition. In Sec.~V, we estimate the luminosity and discuss its relation to gamma-ray bursts. The results are summarized and discussed in Sec.~VI.

Throughout the paper, we use the metric signature \((-+++) \) and geometrized units $c=8\pi G=1$.

\section{Radiation Density in Gravitational Collapse with Phase Transition}

We aim to determine the behavior of radiation density during the gravitational collapse of baryonic matter undergoing a phase transition, which results in the formation of a new type of matter that prevents the formation of a singularity. In this section, we outline the general method for finding the radiation density in this process.

Our analysis begins with a spherically symmetric dynamical spacetime describing a black hole, written in Eddington-Finkelstein coordinates:
\begin{equation} \label{eq:metric}
ds^2 = -\left(1 - \frac{2M(v,r)}{r}\right) dv^2 + 2 dv dr + r^2 d\Omega^2,
\end{equation}
where $M(v,r)$ is the dynamic mass function representing the energy content within a sphere of radius $r$, $d\Omega^2 = d\theta^2 + \sin^2\theta \, d\varphi^2$ is the metric on the unit 2-sphere, and $v$ denotes the Eddington time.

The energy-momentum tensor corresponding to the spacetime \eqref{eq:metric} is a combination of type I and type II matter. Here, type I refers to ordinary matter with $T^0_0 = T^1_1 = -\rho$ and $T^2_2 = T^3_3 = P$, while type II represents radiation flux density $T^1_0 = \sigma$. Thus, applying Einstein’s equations to the dynamical spacetime \eqref{eq:metric} leads to the following relations:
\begin{eqnarray} \label{eq:density}
\sigma &=& \frac{2\dot{M}}{r^2}, \nonumber \\
\rho &=& \frac{2M'}{r^2}, \nonumber \\
P &=& -\frac{M''}{r},
\end{eqnarray}
where the dot and prime denote differentiation with respect to $v$ and $r$, respectively.

To determine the function $M(v,r)$, it is necessary to specify an equation of state in the form $P = P(\rho)$. Moreover, one of the conservation laws for the energy-momentum tensor, $T^{ik}_{;k} = 0$, yields the continuity equation:
\begin{equation} \label{eq:continuous}
\rho' r + 2\rho + 2P = 0.
\end{equation}

In previous studies~\cite{bib:vertogradov2025podu, bib:ali2025jcap, bib:vertogradov2025epjc}, models of gravitational collapse were proposed wherein baryonic matter transitions into a new type of substance with associated radiation emission. This implies that during the formation of the new matter, the total combination of baryonic matter plus radiation is conserved, but their individual components exchange energy, leading to non-conserved parts of the energy-momentum tensor for each component separately. We can express this as a system:
\begin{eqnarray} \label{eq:system}
\rho_r' r &+& 2\rho_r + 2P_r = \beta(v,r) \rho_r, \nonumber \\
\rho_b' r &+& 2\rho_b + 2P_b = -\beta(v,r) \rho_r,
\end{eqnarray}
where $\beta(v,r)$ defines the dimensionless rate of conversion of baryonic matter into a new phase with radiation emission. It is assumed that $\beta > 0$ for all $r$ and $v$, meaning that we exclude the reverse process of radiation converting back into baryonic matter. An important remark is necessary: a dependence of $\beta$ on coordinate quantities would violate general covariance. Therefore, this function must depend on invariant quantities. As shown in~\cite{bib:vertogradov2025epjc}, $\beta$ depends solely on the energy density $\rho_{\text{new}}$ and pressure $P_{\text{new}}$ of the newly formed matter, i.e., $\beta \equiv \beta(\rho_{\text{new}}, P_{\text{new}})$.

Since the metric \eqref{eq:metric} describes a spacetime supported by an anisotropic matter distribution, the barotropic equation of state should be written in the form~\cite{bib:kiselev, bib:husain1998exact}:
\begin{equation} \label{eq:omega}
P = \frac{1}{2} (3\omega + 1) \rho,
\end{equation}
where $\omega$ is the barotropic parameter. Introducing the auxiliary parameter $\alpha$ for computational convenience,
\begin{equation} \label{eq:alpha}
\alpha \equiv \frac{1}{2} (3\omega + 1),
\end{equation}
and comparing with \eqref{eq:omega}, we find that as $\omega$ ranges from $0$ to $1$, $\alpha \in \left[\frac{1}{2}, 2\right]$.

Thus, radiation with $\omega = \frac{1}{3}$ corresponds to $\alpha = 1$, and its equation of state is:
\begin{equation}
P_r = \rho_r.
\end{equation}
Substituting this into the first equation of \eqref{eq:system} and integrating, we obtain the radiation energy density:
\begin{equation} \label{eq:radiation}
\rho_r = \rho_0 \, \exp\left( \int \frac{\beta - 4}{r} \, dr \right).
\end{equation}
Now, substituting \eqref{eq:radiation} into the second equation of \eqref{eq:system} and using the equation of state \eqref{eq:omega}, integration yields:
\begin{equation} \label{eq:baryonic}
\rho_b = r^{-(2\alpha + 2)} \left[ C(v) - \int r^{1+2\alpha} \, \beta \, \rho_0 \, e^{\int \frac{\beta - 4}{r} dr} \, dr \right].
\end{equation}

During the physical process of baryonic matter converting into radiation \eqref{eq:system}, a new type of matter $\rho_{\text{new}}$ is formed:
\begin{equation}
\rho_b + \rho_r = \rho_{\text{new}}.
\end{equation}
Using expressions \eqref{eq:radiation} and \eqref{eq:baryonic}, we can derive the radiation density as a function depending on the new type of matter:
\begin{equation} \label{eq:main}
\rho_r = \frac{\alpha \, \rho_{\text{new}} - P_{\text{new}}}{\alpha - 1}.
\end{equation}
Note that expression \eqref{eq:main} prohibits the value $\alpha = 1$, as this would correspond to a system where radiation transforms into radiation, reducing to the ordinary Einstein equations and leading to the well-known Bonnor–Vaidya solution~\cite{bib:bonor} - a dynamical generalization of the Reissner–Nordström solution.

Our objective is to investigate the energy density \eqref{eq:main} for known regular black hole models, such as those of Dymnikova, Bardeen, and Hayward, in order to understand how the radiation density depends on the specific type of matter that prevents singularity formation.

\section{Matching of Two Solutions: The Kiselev Junction}

The gravitational collapse model can be described by three stages: 1. collapse of baryonic matter; 2. transitional regime, in which  baryonic matter collapses  with transition to a new type of matter accompanied by radiation emission;  3. final stage when baryonic matter completely transforms into a new type of matter.

From the previous section, we know how to compute the radiation density $\rho_r$ depending on the matter into which the barotropic material transitions. However, to estimate the radiation density and avoid its negative values, we must determine the intervals within which the radial coordinate varies. This will also help us estimate the luminosity in subsequent analysis.

We consider radiation in the interval from the critical radius $r_c$, where baryonic matter has already completely transformed into a new type of substance, and therefore, according to formula \eqref{eq:main}, the radiation vanishes $\rho_r(r_c)=0$. To achieve this, we need to match two metrics at this radius: the interior spacetime metric supported by the new type of matter, and the exterior spacetime metric described by a barotropic matter distribution with the equation of state $P=\frac{1}{2}\left(3\omega+1\right) \rho$. Note that the exterior metric represents the Kiselev solution~\cite{bib:kiselev, bib:husain1998exact}.

Consider the Kiselev metric in Eddington--Finkelstein coordinates:
\begin{equation}
ds^2 = -f_K(r) dv^2 + 2 dv dr + r^2 d\Omega^2,
\end{equation}
where
\begin{equation}\label{eq:kiselev}
f_K(r) = 1 - \frac{2 M_K(r)}{r}, \quad
M_K(r) = M_{0k} + \frac{N}{2 r^{3\omega}},
\end{equation}
\begin{equation}
\rho_K(r) = -\frac{3\omega N}{r^{3\omega+3}}, \quad
P_K(r) = \frac{1}{2}\left(3\omega+1\right)\rho_K(r)=\alpha \rho_K(r).
\end{equation}
Here \( \alpha \) is a constant parameter of the equation of state, $M_0$ is a mass of the central object and $N$ is associated with combination of electrical and magnetic charges supported by non-linear electrodynamics~\cite{bib:ijgmmp2026}

Note that for a spherically symmetric metric of the form:
\begin{equation}
ds^2=-f(r)dv^2+2dvdr+r^2d\Omega^2,
\end{equation}
the Einstein equations reduce to second-order linear differential equations:
\begin{eqnarray}
\rho&=&\frac{2M'}{r^2},\nonumber \\
P&=&-\frac{M''}{r}.
\end{eqnarray}
Furthermore, when examining curvature scalars, as well as components of the Ricci and Riemann tensors, it becomes evident that they represent a superposition of the mass function $M(v,r)$ and its first and second derivatives with respect to the radius $M'(r,v)$, $M''(r,v)$:
\begin{eqnarray} \label{eq:curvature}
R &=& \frac{4M' + 2rM''}{r^2}, \nonumber \\
S &=& \frac{8M'^2 + 2r^2 M''^2}{r^4}, \nonumber \\
K &=& \frac{48M^2 - 64rMM' + 32r^2 M'^2 + 16r^2 MM'' - 16r^3 M' M'' + 4r^4 M''^2}{r^6}.
\end{eqnarray}
Thus, for a smooth matching of the Kiselev metric \eqref{eq:kiselev} with the interior solution, it is necessary and sufficient to satisfy the following relations:
\begin{equation}
M_{k}(r_c)=M_{inner}(r_c) \quad,\quad M_k'(r_c)=M_{inner}'(r_c) \quad, \quad M''_k(r_c)=M''_{inner}(r_c).
\end{equation}

From condition (\ref{eq:eos_match}), the critical radius \( r_c \) is determined. Then from (\ref{eq:density_match}), the constant \( N \) is found, and from (\ref{eq:mass_match}), the constant \( M_{0k} \) is obtained.

For regular metrics (Dymnikova, Hayward, Bardeen) at the critical point \( r = r_c \), we require three conditions to be satisfied:
\begin{align}
    M_{\text{reg}}(r_c) &= M_K(r_c), \label{eq:mass_match} \\
    \rho_{\text{reg}}(r_c) &= \rho_K(r_c), \label{eq:density_match} \\
    \left.\frac{P}{\rho}\right|_{\text{reg}}(r_c) &= \alpha. \label{eq:eos_match}
\end{align}
\subsection{Dymnikova Regular Black Hole}

Dymnikova~\cite{bib:dym1} employed the idea of Gliner~\cite{bib:gliner} and Sakharov~\cite{bib:sakharov} that, at critical densities, matter undergoes a phase transition into a vacuum-like medium with a de Sitter core at the center (i.e., the equation of state $P=-\rho$ is realized at the origin).

The Dymnikova solution in Eddington--Finkelstein coordinates takes the form~\cite{bib:shatov2025cpc}:
\begin{align}
    ds^2 = -f(r,v)\, dv^2 + 2\, dv\, dr + r^2 d\Omega^2.
\end{align}
The lapse function $f(r,v)$ is given by
\begin{align} \label{eq:dim}  
    f_d(r,v) = 1 - \frac{2M(v)}{r}\left(1 - e^{-\frac{r^3}{2 M(v) r_0^2}}\right),
\end{align}
where $M(v)$ is the dynamical black hole mass and $r_0$ is the radius of the de Sitter core.

As discussed above, our goal is to identify the region in which baryonic matter completely transforms into Dymnikova-type matter. To this end, we equate the equations of state of baryonic matter and Dymnikova matter, which allows us to determine a critical radius $r_c$ at which the transition occurs. Then, by matching the energy densities in the Dymnikova and Kiselev metrics, we determine the relation between the parameters $r_0$ and $N$. Finally, by equating the mass functions in the Kiselev and Dymnikova metrics, we obtain a relation between the Kiselev black hole mass and the Dymnikova black hole mass, thereby completing the matching procedure.

To proceed, let us write the mass functions for the Kiselev metric:
\begin{equation} \label{eq:kiselevmass}
M_k(v,r)=M_{0k}(v)+\frac{N}{2r^{3\omega}},
\end{equation}
and for the Dymnikova metric:
\begin{equation} \label{eq:dimmass}
M_d(v,r)=M_0(v)\left(1-e^{-\frac{r^3}{2M_0(v)r_0^2}}\right).
\end{equation}

Using the mass function and the Einstein equation $\rho=\frac{2M'}{r^2}$, we obtain the energy density in the Kiselev metric:
\begin{equation} \label{eq:kiselev_density}
\rho_k=-\frac{3\omega N}{r^{3\omega+3}},
\end{equation}
and in the Dymnikova metric:
\begin{equation} \label{eq:dim_density}
\rho_d=\frac{3}{r_0^2}e^{-\frac{r^3}{2M_0(v)r_0^2}}.
\end{equation}

Note that the positivity of energy requires that for the barotropic parameter $\omega \in [0,1]$, the parameter $N$ must be non-positive, i.e., $N\leq 0$. For convenience, we also introduce the notation $\varepsilon = \frac{3}{r_0^2}$.

Using again the Einstein equation relating the mass function and pressure, $P=-\frac{M''}{r}$, we find the pressure in the Kiselev metric:
\begin{equation} \label{eq:kiselev_pressure}
P_k=-\frac{1}{2}\left(3\omega+1\right)\frac{3\omega N}{r^{3\omega+3}}=\frac{1}{2}\left(3\omega+1\right)\rho_k,
\end{equation}
and the pressure in the Dymnikova metric:
\begin{equation} \label{eq:dim_pressure}
P_d=\left(-1+\frac{\varepsilon r^3}{4M_0}\right)\varepsilon e^{-\frac{r^3}{2M_0 r_0^2}}=\left(-1+\frac{\varepsilon r^3}{4M_0}\right)\rho_d.
\end{equation}

However, the equation of state in the form \eqref{eq:dim_pressure} explicitly depends on the radial coordinate $r$, which at first glance violates general covariance. Therefore, it is necessary to express the pressure $P_d$ purely as a function of the energy density $\rho_d$. This can be done in the form
\begin{equation} \label{eq:dim_eos}
P_d=\left(-1-\frac{3}{2}\ln \left|\frac{\rho_d}{\varepsilon}\right|\right)\rho_d,
\end{equation}
which represents a generalization of the Hagedorn equation of state~\cite{bib:hagedorn, bib:malafarina, bib:thurco_hagedorn, bib:vertogradov2025cqg}.

From the equality of pressures \eqref{eq:kiselev_pressure} and \eqref{eq:dim_pressure}, we determine the critical radius $r=r_c$, i.e., the radius at which baryonic matter and Dymnikova matter \eqref{eq:dim_eos} share the same equation of state. For convenience, we introduce
\begin{equation}
\alpha=\frac{1}{2}\left(3\omega+1\right), \quad k=\frac{\varepsilon}{4M_0}.
\end{equation}

We then obtain the critical radius $r_c$:
\begin{equation}
k r_c^3 - 1 = \alpha \quad \Rightarrow \quad
r_c = \left( \frac{\alpha + 1}{k} \right)^{1/3}.
\end{equation}

Next, we equate the energy densities \eqref{eq:kiselev_density} and \eqref{eq:dim_density} in order to determine the parameter $N$ as a function of $r_c$ and $r_0$, i.e., $N \equiv N(r_c,r_0)$:
\begin{equation}
\rho_d(r_c) = \varepsilon e^{-\frac{r_c^3}{2 M_0 r_0^2}} = \varepsilon e^{-\frac{2}{3} k r_c^3}
           = \varepsilon e^{-\frac{2}{3}(\alpha + 1)},
\end{equation}
\begin{equation}
\rho_K(r_c) = \frac{(1 - 2\alpha) N}{r_c^{2\alpha + 2}}.
\end{equation}
Hence,
\begin{equation}
N(r_c,r_0) = \frac{\varepsilon r_c^{2\alpha + 2}}{(1 - 2\alpha)} e^{-\frac{2}{3}(\alpha + 1)}.
\end{equation}

Finally, we equate the mass functions \eqref{eq:kiselevmass} and \eqref{eq:dimmass} to express the Kiselev black hole mass $M_{0k}$ as a function of $r_c$, $r_0$, and $M_0$, i.e., $M_{0k}\equiv M_{0k}(r_c,r_0,M_0)$:
\begin{equation}
M_d(r_c) = M_0 \left( 1 - e^{-\frac{r_c^3}{2 M_0 r_0^2}} \right) = M_0 \left( 1 - e^{-\frac{2}{3}(\alpha + 1)} \right).
\end{equation}
The Kiselev mass function at $r_c$ is
\begin{equation}
M_K(r_c) = M_{0k} + \frac{N}{2 r_c^{2\alpha - 1}}.
\end{equation}
Equating the two expressions, we obtain
\begin{equation}
M_{0k} = M_d(r_c) - \frac{N}{2 r_c^{2\alpha - 1}}.
\end{equation}

For the critical radius $r_c$ to be real, it is necessary that $\alpha + 1 > 0$, i.e., $\alpha > -1$. However, as discussed above, we are interested in the physically relevant interval $\omega \in [0,1]$, which corresponds to $\alpha \in \left[\frac{1}{2},2\right]$. This guarantees the reality of $r_c$ within the considered model.

Additionally, we emphasize that the matching procedure constructed here ensures a smooth transition between baryonic matter described by the Kiselev solution and the regular core described by the Dymnikova solution. This provides a physically consistent framework for modeling gravitational collapse leading to regular black hole formation with a de Sitter interior.

\subsection{Hayward Regular Black Hole}

Considering the formation and evaporation of regular black holes, Hayward proposed a minimal model describing a dynamical regular black hole of the form~\cite{bib:hay}:
\begin{equation} \label{eq:hay}
f_h(r,v) = 1 - \frac{2 M_{0h}(v) r^2}{r^3 + 2 M_{0h}(v) L^2},
\end{equation}
where $M_{0h}(v)$ is the black hole mass and $L$ is the regularization parameter.

As in the previous section, we proceed to match the Kiselev metric \eqref{eq:kiselev} with the Hayward metric \eqref{eq:hay}, following the same algorithm as for the Dymnikova regular black hole.

To this end, we need the mass function
\begin{equation} \label{eq:hay_mass}
M_h(v,r)=\frac{M_{0h}(v)\, r^3}{r^3+2M_{0h}(v)L^2},
\end{equation}
the energy density
\begin{equation} \label{eq:hay_density}
\rho_h=\frac{12M_{0h}^2(v)L^2}{\left(r^3+2M_{0h}(v)L^2\right)^2},
\end{equation}
and the pressure
\begin{equation} \label{eq:hay_pressure}
P_h=\left(-1+\frac{3r^3}{r^3+2M_{0h}(v)L^2}\right)\frac{12M_{0h}^2(v)L^2}{\left(r^3+2M_{0h}(v)L^2\right)^2}
=\left(-1+\frac{3r^3}{r^3+2M_{0h}(v)L^2}\right)\rho_h.
\end{equation}

As in the case of the Dymnikova regular black hole, the equation of state in the form \eqref{eq:hay_pressure} is not manifestly generally covariant, since it explicitly depends on the radial coordinate $r$. Expressing the pressure solely as a function of the energy density leads to a generalized polytropic equation of state~\cite{bib:yi2026epjc}:
\begin{equation} \label{eq:hay_eos}
P_h = 2\rho_h - \sqrt{3}\,L\,\rho_h^{3/2}.
\end{equation}

Now, assuming that at some radius $r=r_c$ the equation of state of barotropic matter coincides with the Hayward equation of state \eqref{eq:hay_pressure}, we find the critical radius from
\begin{equation}
\frac{2(r_c^3 - M_{0h} L^2)}{r_c^3 + 2 M_{0h} L^2} = \alpha.
\end{equation}
Solving with respect to $r_c^3$, we obtain
\begin{equation}
r_c^3 = \frac{2 M_{0h} L^2 (\alpha + 1)}{2 - \alpha}, \quad \alpha \neq 2.
\end{equation}

Next, we compare the energy densities at the critical radius in the Kiselev metric \eqref{eq:kiselev_density} and the Hayward spacetime \eqref{eq:hay_density}:
\begin{equation}
\rho_h(r_c) = \frac{12 M_{0h}^2 L^2}{(r_c^3 + 2 M_{0h} L^2)^2}.
\end{equation}
Using the expression for $r_c^3$, we find
\begin{equation}
r_c^3 + 2 M_{0h} L^2 = \frac{6 M_{0h} L^2}{2 - \alpha}, \quad
\rho_h(r_c) = \frac{(2 - \alpha)^2}{3 L^2}.
\end{equation}

Equating this to the Kiselev density $\rho_k(r_c)$, we obtain
\begin{equation}
\frac{(1 - 2\alpha) N}{r_c^{2\alpha + 2}} = \frac{(2 - \alpha)^2}{3 L^2}.
\end{equation}
Hence,
\begin{equation}
N = \frac{(2 - \alpha)^2 r_c^{2\alpha + 2}}{3 L^2 (1 - 2\alpha)}.
\end{equation}

In complete analogy, we now express the mass $M_{0k}$ using \eqref{eq:kiselevmass}. The Hayward mass function at $r_c$ is
\begin{equation}
M_h(r_c) = \frac{M_{0h} r_c^3}{r_c^3 + 2 M_{0h} L^2} = \frac{M_{0h} (\alpha + 1)}{3}.
\end{equation}
Equating this with $M_K(r_c)$, we obtain
\begin{equation}
M_{0k} = M_h(r_c) - \frac{N}{2 r_c^{2\alpha - 1}}.
\end{equation}

It is important to note that $\alpha \neq 2$. Furthermore, for the positivity of $M_{0k}$ with $M_{0h} > 0$, additional constraints on $\alpha$ arise. In particular, depending on the sign of $(1 - 2\alpha)$, one typically requires $\alpha < \frac{4}{5}$ (and, in particular, $\alpha > \frac{1}{2}$ ensures consistency with the sign of $N$), which is compatible with the physically relevant range of parameters in the model.

We emphasize that this matching procedure provides a consistent transition between a Kiselev-type exterior supported by barotropic matter and a regular Hayward core, thus extending the construction of dynamical regular black hole spacetimes with physically motivated equations of 
state.

\subsection{Bardeen Regular Black Hole}

In the 1960s, Bardeen obtained the first solution of the Einstein equations describing a regular black hole~\cite{bib:bardeen}. The lapse function $f_B(r,v)$ for this metric is given by~\cite{bib:angel2025podu}:
\begin{equation} \label{eq:bardeen}
f_B(r,v) = 1 - \frac{2 M_{0b}(v) r^2}{(r^2 + g^2)^{3/2}},
\end{equation}
where $M_{0b}(v)$ is the black hole mass and $g$ is a parameter associated with a magnetic charge supported by nonlinear electrodynamics~\cite{bib:garsia}.

In this subsection, we follow exactly the same matching procedure as in the two previous cases. To ensure a smooth matching, we require the mass function
\begin{equation} \label{eq:bardeen_mass}
M_b(v,r)=\frac{M_{0b}(v)\,r^3}{(r^2+g^2)^{3/2}},
\end{equation}
the energy density
\begin{equation} \label{eq:bardeen_density}
\rho_b = \frac{6 M_{0b}(v) g^2}{(r^2 + g^2)^{5/2}}, 
\end{equation}
and the pressure
\begin{equation} \label{eq:bardeen_pressure}
P_b= -\frac{6M_{0b}(v) g^2}{(r^2 + g^2)^{5/2}} \left( 1 - \frac{5 r^2}{2(r^2 + g^2)} \right).
\end{equation}

As in the previous cases, the equation of state expressed in this form explicitly depends on the radial coordinate $r$, and thus its general covariance is not manifest. However, it was shown in~\cite{bib:yi2026epjc} that it can be rewritten in a form analogous to the equation of state \eqref{eq:hay_eos}. 

Our goal is to determine the radius $r=r_c$ at which Bardeen matter transitions into barotropic matter. To this end, we define an effective equation-of-state parameter:
\begin{equation}
\alpha_b(r) = \frac{P_b}{\rho_b} = -1 + \frac{5 r^2}{2(r^2 + g^2)}.
\end{equation}
Equating this effective coefficient to the barotropic one, we obtain the critical radius $r=r_c$.

From the condition $\alpha_b(r_c) = \alpha$, we have:
\begin{equation}
-1 + \frac{5 r_c^2}{2(r_c^2 + g^2)} = \alpha.
\end{equation}
Solving with respect to $r_c^2$, we find:
\begin{equation}
r_c^2 = \frac{2 (\alpha + 1) g^2}{3 - 2\alpha}, \quad \alpha \neq \frac{3}{2}.
\end{equation}

Next, using the definition of the critical radius, we determine how the parameter $N$ depends on $r_c$ and the magnetic charge $g$:
\begin{equation}
\rho_b(r_c) = \frac{6 M_{0b} g^2}{(r_c^2 + g^2)^{5/2}}.
\end{equation}
Using the expression for $r_c^2$, we obtain:
\begin{equation}
r_c^2 + g^2 = \frac{5 g^2}{3 - 2\alpha}, \quad
\rho_b(r_c) = \frac{6 M_{0b} (3 - 2\alpha)^{5/2}}{5^{5/2} g^3}.
\end{equation}

Equating this to $\rho_K(r_c)$, we obtain:
\begin{equation}
\frac{(1 - 2\alpha) N}{r_c^{2\alpha + 2}} = \frac{6 M_{0b} (3 - 2\alpha)^{5/2}}{5^{5/2} g^3}.
\end{equation}
Hence,
\begin{equation}
N = \frac{6 M_{0b} (3 - 2\alpha)^{5/2}}{5^{5/2} g^3 (1 - 2\alpha)} r_c^{2\alpha + 2}.
\end{equation}

We now express the Kiselev mass parameter $M_{0k}$ in terms of the Bardeen metric parameters.

The Bardeen mass function at $r_c$ is
\begin{equation}
M_b(r_c) = \frac{M_{0b} r_c^3}{(r_c^2 + g^2)^{3/2}}.
\end{equation}
Substituting the expressions for $r_c^2$ and $r_c^2 + g^2$, we obtain:
\begin{equation}
r_c^3 = \left( \frac{2 (\alpha + 1) g^2}{3 - 2\alpha} \right)^{3/2}, \quad
(r_c^2 + g^2)^{3/2} = \left( \frac{5 g^2}{3 - 2\alpha} \right)^{3/2}.
\end{equation}
Thus,
\begin{equation}
M_b(r_c) = M_{0b} \left( \frac{2(\alpha+1)}{5} \right)^{3/2}.
\end{equation}

Next, we evaluate the second term in $M_K(r_c)$:
\begin{equation}
\frac{N}{2 r_c^{2\alpha - 1}} = \frac{N}{2} r_c^{1 - 2\alpha}.
\end{equation}
Substituting the expression for $N$, we obtain
\begin{equation}
\frac{N}{2 r_c^{2\alpha - 1}} = \frac{3 M_{0b} g^2 \, r_c^{3}}{(1 - 2\alpha) \left( r_c^2 + g^2 \right)^{5/2}}.
\end{equation}
Substituting $r_c^2$ and $r_c^2 + g^2$, we find
\begin{equation}
\frac{N}{2r_c^{2\alpha-1}}
=
\frac{3M_{0b}}{1-2\alpha}
\frac{[2(\alpha+1)]^{3/2}(3-2\alpha)}
{5^{5/2}}.
\end{equation}
Finally, from the condition $M_K(r_c) = M_b(r_c)$, we obtain:
\begin{equation}
M_{0k} = M_b(r_c) - \frac{N}{2 r_c^{2\alpha - 1}}.
\end{equation}

We emphasize that this construction provides a consistent matching between the Kiselev exterior supported by barotropic matter and the Bardeen regular core, thus extending the class of dynamically matched regular black hole solutions with physically motivated matter sources.

\section{Radiation Density During the Formation of a Regular Black Hole}

The radiation density arising during the phase transition of baryonic matter into a new form of matter preventing singularity formation depends on the size of the collapsing star. As noted above, at the initial stage of the collapse no phase transition occurs, and therefore such radiation is absent. At the critical radius $r=r_c$, when all barotropic matter has completely transformed into the new type of matter, the radiation also vanishes. This follows directly from the previously obtained expression for the radiation density:
\begin{equation} \label{eq:radden}
\rho_r=\frac{\alpha \rho_{\text{new}}-P_{\text{new}}}{\alpha-1}.
\end{equation}
At the critical point, the equations of state of the new matter and baryonic matter coincide, implying that the numerator on the right-hand side vanishes. Therefore, it is necessary to investigate the behavior of the radiation density in the region where the phase transition begins and up to the region $r=r_c$, where the transition is completed and radiation ceases.

Let us first obtain the radiation density for the Dymnikova metric. The energy density and pressure are given by
\begin{equation}
\rho_d(r) = \varepsilon e^{-\frac{r^3}{2 M r_0^2}}, \quad
P_d(r) = \rho_d(r)\left( \frac{3 r^3}{4 M r_0^2} - 1 \right).
\end{equation}
For simplicity, in these calculations we set $M_0=1$. At the critical point, the equation of state takes the form
\begin{equation}
P_{\text{new}}=\alpha \rho_{\text{new}}.
\end{equation}
Then,
\begin{equation}
\rho_r=\frac{\alpha \rho_{\text{new}}-\alpha \rho_{\text{new}}}{\alpha-1} = 0.
\label{eq:rho_r_zero}
\end{equation}
Thus, independently of the specific metric, the radiation density vanishes at the critical point.

\begin{figure}
    \centering
    \includegraphics[width=0.6\linewidth]{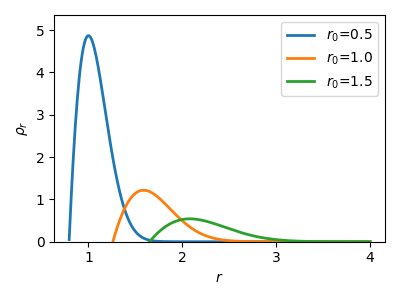}
    \caption{Radiation density $\rho_r(r)$ in the Dymnikova metric for different values of $r_0$.}
    \label{fig:dym}
\end{figure}

We now analyze the phase transition region. Using the expression above, we study the behavior of the radiation density $\rho_r(r)$ for different admissible values of the regularization parameters. As established in the previous sections for regular black holes, we fix the mass function normalization and consider $\alpha = 1/2$. In Fig.~\ref{fig:dym} we present the profile of $\rho_r(r)$ for three different values of the Dymnikova parameter $r_0$, ranging from a minimal value $r_0=0.5$ to an extremal value $r_0=1.5$. The figure shows that the maximum radiation density corresponds to the smallest value of $r_0=0.5$. As $r_0$ increases toward its extremal value, the profile becomes progressively flatter. In the limit $r \to \infty$, the radiation density $\rho_r$ tends to zero.

\begin{figure}
    \centering
    \includegraphics[width=0.6\linewidth]{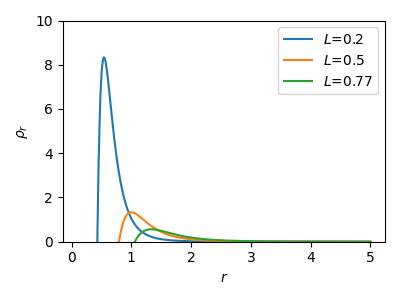}
    \caption{Radiation density $\rho_r(r)$ in the Hayward metric for different values of $L$.}
    \label{fig:hay}
\end{figure}

In Fig.~\ref{fig:hay}, we show the radiation density in the Hayward metric. Notably, for the minimal value of the regularization parameter $L=0.2$, the radiation density reaches its maximum. Moreover, this maximal value of $\rho_r$ in the Hayward case exceeds the corresponding maximal radiation density in the Dymnikova case.

\begin{figure}
    \centering
    \includegraphics[width=0.6\linewidth]{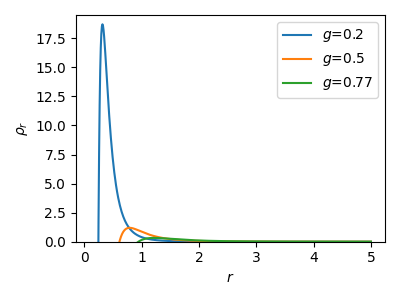}
    \caption{Radiation density $\rho_r(r)$ in the Bardeen metric for different values of $g$.}
    \label{fig:bar}
\end{figure}

Finally, in Fig.~\ref{fig:bar}, we present the radiation density for the Bardeen metric. A similar behavior is observed: the radiation density $\rho_r$ attains its maximum for the minimal value of the regularization parameter $g=0.2$. Furthermore, for this value, the radiation density exceeds that of the Hayward case with the same regularization parameter $L=0.2$.

Thus, we conclude that at the onset of the phase transition, the radiation density increases. This is physically explained by the fact that as the object becomes denser, the conversion rate of baryonic matter into the new type of matter accelerates. Eventually, this transition rate reaches a maximum, at which point the radiation density also attains its peak value. Afterward, both the transition rate and the radiation density decrease until the baryonic matter is completely converted into the new phase. At the critical radius $r=r_c$, the transition rate vanishes, and consequently, the radiation density also becomes zero.
\section{Luminosity during the phase transition}

The transition from baryonic matter to an exotic phase is accompanied by the release of energy. In what follows, we estimate the corresponding bolometric luminosity. The quantity obtained in this way should not be identified with an optical luminosity, since the spectral distribution of the emitted radiation is not fixed by the gravitational field equations alone.

Let \(\rho_{\rm phys}\) be the physical energy density of the outgoing radiation in cgs units. For an approximately radial null flux one has
\begin{equation}
F=c\rho_{\rm phys}.
\end{equation}
The luminosity measured through a sphere of radius \(r\) is then
\begin{equation}
L=4\pi r^2F,
\end{equation}
or
\begin{equation}
L=4\pi r^2c\rho_{\rm phys}.
\end{equation}
Restoring cgs units in the expression obtained from the field equations gives
\begin{equation}
\rho_{\rm phys}={c^4\over8\pi G}\rho_r .
\end{equation}
Therefore the dominant dimensional factor in the luminosity is
\begin{equation}
L_0={c^5\over G},
\end{equation}
which is numerically equal to
\begin{equation}
L_0=3.63\times10^{59}\ {\rm erg\,s^{-1}}.
\end{equation}
Observed gamma-ray bursts may reach luminosities of order
\begin{equation}
10^{55}\ {\rm erg\,s^{-1}}.
\end{equation}
Thus, the remaining dimensionless factor \(r^2\rho_r/2\) must be of order
\begin{equation}
10^{-4}
\end{equation}
in order to be compatible with this scale.

The radiation density \(\rho_r\), given in \eqref{eq:radden}, depends both on the initial stellar matter and on the final matter produced by the phase transition. Using \eqref{eq:radden}, we obtain
\begin{equation}
\frac{r^2}{2}\left(\rho_{new}+\frac{\rho_{new}-P_{new}}{\alpha-1}\right) \lesssim 10^{-4}.
\end{equation}
Equivalently, using the Einstein equations, this condition can be written as
\begin{equation}
M_{new}'+\frac{M_{new}'+\frac{1}{2}M_{new}''r}{\alpha-1}\lesssim 10^{-4}.
\end{equation}
If \(\alpha>1\) and the dominant energy condition is satisfied, namely
\begin{equation}
\rho_{new}-P_{new}\geq 0,
\end{equation}
then the radiation contribution is bounded from below by $M'_{new}$. Hence, in order to understand what restrictions the observed gamma-ray burst luminosities impose on the model, one has to analyze the allowed values of \(M'\).

In the present paper, we consider asymptotically flat regular black holes with two horizons. These geometries are described by two parameters: the black-hole mass \(M_0\) and the regularization parameter, denoted here by \(\xi\). We also consider a stellar model whose matter satisfies a barotropic equation of state with \(\alpha>1\). It is important to stress that this does not necessarily imply a violation of the dominant energy condition. As discussed above, for the averaged pressure
\begin{equation}
\bar{P}=\frac{1}{3}\left(P_r+2P_t\right)=\omega \rho
\end{equation}
there is a relation between the true barotropic parameter \(\omega\) and the anisotropic barotropic coefficient \(\alpha\),
\begin{equation}
\alpha =\frac{1}{2}\left(3\omega+1\right).
\end{equation}
Therefore, if \(\omega \in [0,1]\), then
\begin{equation}
\alpha \in \left[\frac{1}{2},2\right].
\end{equation}

During gravitational collapse, horizons are formed. Phase transitions occurring under the event horizon are no longer accessible to an external observer. Consequently, most of the emitted radiation is hidden as well, since it cannot reach infinity. For this reason, the relevant estimate must be made near the outer event horizon \(r=r_{+}\). The location of the horizons depends on the black-hole mass \(M_0\) and on the regularization parameter \(\xi\). As shown in Refs.~\cite{bib:vertogradov2025plb,bib:ali2024plb}, if the weak energy condition is satisfied,
\begin{equation}
\rho\geq 0 \Rightarrow M'\geq 0,
\end{equation}
then increasing the regularization parameter \(\xi\) makes the outer horizon shrink and the inner horizon expand. At a certain critical value of \(\xi\), the two horizons merge and an extremal black hole is formed.

For asymptotically flat regular black holes, the mass function must obey
\begin{equation}
M(r)|_{\xi=0}=M_0,
\end{equation}
so that in the absence of the regularization parameter one recovers the Schwarzschild solution. Another important property is that, at the extremal horizon, one has
\begin{equation}
1-2M'=0 \quad \text{for an extremal black hole}.
\end{equation}
Combining these properties, we conclude that
\begin{equation}
M'\in [0, \frac{1}{2}].
\end{equation}
Therefore, in order to satisfy the condition
\begin{equation}
\frac{r^2\rho_r}{2}\lesssim 10^{-4},
\end{equation}
the regularization parameter must be small. In other words, the geometry must represent only a minimal deviation from the Schwarzschild metric. If we introduce the parametrization
\begin{equation}
\xi=\eta M_0,
\end{equation}
then the regularization scale \(\xi\) can be discussed in terms of the dimensionless parameter \(\eta\).

Table~\ref{tab:eta-luminosity-bound} shows the values of \(\eta\) obtained for the Dymnikova, Hayward, and Bardeen metrics. The condition imposed in the table is the saturation of the estimate
\begin{equation}
M'(r_+)=10^{-4},
\end{equation}
where the derivative is evaluated at the outer event horizon. This choice is natural because the radiation produced below the event horizon cannot reach a distant observer.

It follows from Table~\ref{tab:eta-luminosity-bound} that the Dymnikova model requires a regularization parameter of order \(10^{-1}\), whereas the Hayward and Bardeen models require parameters of order \(10^{-2}\). In all three cases, the outer horizon remains very close to the Schwarzschild value \(r=2M_0\). Therefore, the luminosity constraint mainly restricts the strength of the regularization sector, while producing only a small displacement of the outer horizon.

\begin{table}[t]
\centering
\caption{
Values of the dimensionless regularization parameter \(\eta\) required for
\(M'(r_+)=10^{-4}\), together with the corresponding position of the outer
event horizon. The mass is fixed as \(M_0=10M_\odot\), where \(M_0\) is written
in geometrized units. For the Dymnikova metric we use
\(r_0=\eta M_0\), for the Hayward metric \(l=\eta M_0\), and for the Bardeen
metric \(g=\eta M_0\). The derivative \(M'\) is evaluated at \(r=r_+\).
}
\label{tab:eta-luminosity-bound}
\begin{tabular}{c c c c}
\hline
Model & \(M(r)\) & \(\eta\) & \(r_+\) \\
\hline
Dymnikova &
\(M_0\left(1-e^{-{r^3\over2M_0r_0^2}}\right)\) &
\(5.747\times10^{-1}\) &
\(1.999989M_0=29.532\ {\rm km}\)
\\
Hayward &
\({M_0r^3\over r^3+2M_0l^2}\) &
\(1.633\times10^{-2}\) &
\(1.999867M_0=29.531\ {\rm km}\)
\\
Bardeen &
\({M_0r^3\over\left(r^2+g^2\right)^{3/2}}\) &
\(1.633\times10^{-2}\) &
\(1.999800M_0=29.530\ {\rm km}\)
\\
\hline
\end{tabular}
\end{table}

Thus, within the present order-of-magnitude estimate, the considered regular black-hole models can be compatible with gamma-ray burst luminosities of order
\begin{equation}
10^{55}\ {\rm erg\,s^{-1}}.
\end{equation}
This requires the regularization sector to give only a weak contribution near the outer horizon. As seen from Table~\ref{tab:eta-luminosity-bound}, the resulting horizon radius differs only slightly from the Schwarzschild value. This is also qualitatively consistent with the observational fact that black-hole shadow measurements allow only moderate deviations from the Schwarzschildshadow size. Therefore, the same small-deviation regime that keeps the luminosity estimate under control also keeps the near-horizon geometry close to the standard black-hole geometry.

\section{Conclusion}

In this paper, we have investigated a possible radiative mechanism associated with the formation of regular black holes during gravitational collapse. The main physical assumption was that ordinary baryonic matter cannot by itself support a nonsingular black-hole core. Therefore, if a regular black hole is formed, a new matter violating the strong energy condition must be generated dynamically during the collapse. Such a transition can release energy, and part of this energy may escape in the form of electromagnetic radiation if the relevant region is not yet hidden from a distant observer.

In our model the matter source was decomposed into a barotropic component, a radiative component, and the final exotic matter . From the conservation of the total energy-momentum tensor, we obtained the radiation density generated during the phase transition,
\begin{equation}
\rho_r=\frac{\alpha\rho_{\rm new}-P_{\rm new}}{\alpha-1}.
\end{equation}
This expression relates the emitted radiation directly to the density and pressure of the matter supporting the final regular black hole. Thus, the radiation is not introduced as an external phenomenological source, but follows from the properties of the matter responsible for removing the singularity.

We applied this construction to three famous models: the Dymnikova, Hayward, and Bardeen regular black holes. For each model , we matched the regular interior with a Kiselev exterior supported by anisotropic barotropic matter. This procedure allowed us to determine the critical radius at which the equation of state of the exterior matter coincides with that of the exotic matter. At this radius, the transition is completed and the radiation density vanishes. Hence, the radiation is produced only in the intermediate region where baryonic matter is converted into the new phase.

The analysis of the radiation density profiles shows that the emitted radiation depends on the regularization scale. This is physically natural: a more compact regular core corresponds to a stronger gravitational field region, where the conversion of baryonic matter into a nonsingular phase can release more energy.

We also estimated the bolometric luminosity associated with this process. In order to be compatible with observed gamma-ray burst , the regularization effects near the outer horizon must be small. For the Dymnikova, Hayward, and Bardeen models, this means that the corresponding regular black holes differ only weakly from the Schwarzschild black hole. This conclusion is consistent with current black hole shadow observations, which allow only limited deviations from the standard black hole shadow size~\cite{bib:eht1,bib:eht2}.

The importance of this result is twofold. First, it shows that the formation of a regular black hole may be accompanied by an observable electromagnetic signal. Therefore, the formation process itself can contain information about the internal structure of the final compact object, which becomes inaccessible after horizon formation. Second, the luminosity of the emitted radiation can constrain the regularization parameters of the final geometry. This provides a possible bridge between high-energy transient observations and the problem of distinguishing singular black holes from regular ones.

The present model should be regarded as a first step to be followed by futher studies. We  have estimated the luminosity from the radiation density, but did not construct a detailed spectrum or a complete model of the prompt emission. Future work should therefore include a more realistic dynamical description of the collapse, including the time dependence of the transition process and the question of whether the emitted radiation can escape before the horizon completely hides the transition region. Another important direction is the extension of the present construction to rotating regular black holes, since astrophysical collapse is expected to produce rotating compact objects.

\section*{Acknowledgements}
The authors acknowledge the support of the National Natural Science Foundation of China under Grant No.  12075059, as well as the start-up fund of USTC.

\bibliographystyle{apsrev4-1}
\bibliography{ref}
\end{document}